\documentclass[a4paper,11pt]{article}
\usepackage{pos}

\usepackage[utf8]{inputenc}
\usepackage{amsmath}
\usepackage{mathrsfs}
\usepackage{graphicx}
\usepackage{bbold}
\usepackage[dvipsnames]{xcolor}
\usepackage{braket}
\usepackage{bbm}

\numberwithin{equation}{section}

\title{Phase diagram of 4D SU(3) Yang-Mills theory at $\theta=\pi$ via imaginary theta simulations}
\ShortTitle{Phase diagram of 4D SU(3) YM theory via imaginary theta}

\author*[a,b]{Akira Matsumoto}
\author*[c,d]{Mitsuaki Hirasawa}
\author[e,f]{Jun Nishimura}
\author[g]{Atis Yosprakob}

\affiliation[a]{
Graduate School of Science, Osaka Metropolitan University,\\
3-3-138 Sugimoto, Sumiyoshi-ku, Osaka 558-8585, Japan}

\affiliation[b]{
RIKEN Center for Interdisciplinary Theoretical and Mathematical Sciences (iTHEMS), RIKEN,\\
2-1 Hirosawa, Wako, Saitama 351-0198, Japan}

\affiliation[c]{
Department of Physics, University of Milano-Bicocca,\\
Piazza della Scienza 3, I-20126 Milano, Italy}

\affiliation[d]{
Istituto Nazionale di Fisica Nucleare (INFN), Sezione di Milano-Bicocca,\\
Piazza della Scienza 3, I-20126 Milano, Italy}

\affiliation[e]{
Graduate Institute for Advanced Studies, SOKENDAI,\\
1-1 Oho, Tsukuba, Ibaraki 305-0801 Japan}

\affiliation[f]{
KEK Theory Center, High Energy Accelerator Research Organization (KEK),\\
1-1 Oho, Tsukuba, Ibaraki 305-0801, Japan}

\affiliation[g]{
Yukawa Institute for Theoretical Physics, Kyoto University,\\
Kitashirakawa Oiwakecho, Sakyo-ku, Kyoto 606-8502 Japan}

\emailAdd{akira.matsumoto(at)omu.ac.jp}
\emailAdd{mitsuaki.hirasawa(at)mib.infn.it}
\emailAdd{jnishi(at)post.kek.jp}
\emailAdd{yosprakob(at)yukawa.kyoto-u.ac.jp}

\abstract{
It has been speculated that the CP symmetry of 4D SU(3) Yang-Mills theory at $\theta=\pi$ is spontaneously broken in the confined phase, 
and it is recovered precisely at the deconfining temperature.
The direct simulation of the theory at $\theta=\pi$ is, however, difficult due to the sign problem.
We therefore simulate the theory with an imaginary theta parameter and perform analytic continuation to the real theta to explore the phase diagram.
We implement the stout smearing technique in the hybrid Monte Carlo simulation to recover the topological property of the gauge field.
The smearing-time dependence of the observable is investigated using the reweighting method with respect to the smearing step parameters, 
and a clear scaling behavior is observed.
The order parameter of the CP symmetry is then computed in the scaling region to detect symmetry breaking.
We report preliminary results on the expected CP breaking and restoration temperature.}

\FullConference{The 42nd International Symposium on Lattice Field Theory (LATTICE2025)\\
2-8 November 2025\\
Tata Institute of Fundamental Research, Mumbai, India\\}

\begin{document}

%%% preprint number
\begin{flushright}
    YITP-26-22, KEK-TH-2813, RIKEN-iTHEMS-Report-26
\end{flushright}

\maketitle

%%%%%%%%%%%%%%%%%%%%%%%%%%%%%%%%%%%%%%%%%%%%%%%%%%%%%%%%%%%%%%%%%%%%%%%%%%%%%%%%
\section{Introduction}
%%%%%%%%%%%%%%%%%%%%%%%%%%%%%%%%%%%%%%%%%%%%%%%%%%%%%%%%%%%%%%%%%%%%%%%%%%%%%%%%

Four-dimensional SU($N$) Yang-Mills theory admits the addition of the topological $\theta$ term to the action 
without violating Lorentz invariance, SU($N$) gauge symmetry, and reormalizability.
Of particular interest in the model is the CP symmetry at $\theta=\pi$.
%For generic $\theta \neq 0$, CP symmetry is explicitly broken.
%Nevertheless, $\theta=\pi$ is a distinguished point: owing to the $2\pi$ periodicity in $\theta$, the theory is CP-invariant there.
The CP symmetry is widely expected to be spontaneously broken at low temperature, 
while it is known to be restored at sufficiently high temperature \cite{Gross:1980br, Weiss:1980rj}.
Recent analyses based on the higher-form symmetries predict---via 't~Hooft anomaly matching---that, 
unless the theory becomes gapless, either the CP or $\mathbb{Z}_N$ center symmetry must be spontaneously broken~\cite{Gaiotto:2017yup}.
This implies the following ordering between the two transitions: 
\begin{equation}
    T_{\mathrm{CP}} \geq T_{\mathrm{dec}}(\pi) \ ,
    \label{inequality-CP-dec}
\end{equation}
where $T_{\mathrm{CP}}$ is the temperature at which CP at $\theta=\pi$ is restored, 
and $T_{\mathrm{dec}}(\pi)$ denotes the deconfining temperature at $\theta=\pi$.

In the large-$N$ limit, these transitions are known to coincide, 
$T_{\mathrm{CP}} = T_{\mathrm{dec}}(\pi)$~\cite{Witten:1980sp, Witten:1998uka, Bigazzi:2015bna}.
In contrast, for $N=2$ one expects $T_{\mathrm{CP}} > T_{\mathrm{dec}}(\pi)$, 
based on studies of SU($N$) supersymmetric Yang-Mills theory deformed by a gaugino mass and compactified on $S^1$, 
where the radius of $S^1$ (with periodic boundary conditions) is treated as an analog of the inverse temperature~\cite{Chen:2020syd}.
This argument requires the gaugino mass to be sufficiently small for supersymmetry-based reasoning to remain trustworthy, 
even though the theory reduces to pure Yang-Mills theory in the infinite mass limit.
Moreover, the $S^1$ radius with periodic boundary conditions is not literally an inverse temperature, 
since a thermal interpretation would require anti-periodic boundary conditions for the gaugino.
In our previous study~\cite{Hirasawa:2024fjt}, we carried out first-principles calculations for the imaginary $\theta$ 
and analytically continued the results to real $\theta$ at fixed lattice spacing.
A dynamical implementation of stout smearing makes this analytic continuation feasible.
Our findings are also consistent with the expectation stated above.
In this work, we extend our study to the $N=3$ case, 
in which the order of the deconfining transition is speculated to be different from that in the $N=2$ case but the same as that in the large-$N$ limit.

%%%%%%%%%%%%%%%%%%%%%%%%%%%%%%%%%%%%%%%%%%%%%%%%%%%%%%%%%%%%%%%%%%%%%%%%%%%%%%%%
\section{4D SU(\texorpdfstring{$N$}{N}) Yang-Mills theories with the \texorpdfstring{$\theta$}{theta} term}
%%%%%%%%%%%%%%%%%%%%%%%%%%%%%%%%%%%%%%%%%%%%%%%%%%%%%%%%%%%%%%%%%%%%%%%%%%%%%%%%

In the Euclidean space, the partition function of 4D SU($N$) Yang-Mills theory with the $\theta$ term is defined by 
\begin{equation}
    Z_\theta = \int \mathcal{D} A\, e^{-S_g + i\theta Q},
\end{equation}
where $S_g$ is the action of the pure Yang-Mills theory and $Q$ is the topological charge, 
\begin{equation}
    Q = \frac{1}{32\pi^2} \int d^4x\, \epsilon_{\mu\nu\rho\sigma} \mathrm{Tr}\left( F_{\mu\nu} F_{\rho\sigma} \right).
\end{equation}
The topological charge takes integer values on a compact manifold, 
and thus the theory has a periodicity of the $\theta$ parameter as $\theta \to \theta + 2\pi$.
Since the topological charge $Q$ is a CP-odd quantity, the nonzero $\theta$ term breaks the CP symmetry.
However, in the specific case, $\theta=\pi$ (up to $2 \pi n$ shift), the CP symmetry is kept intact thanks to the periodicity of $\theta$.

In Monte Carlo simulation, the topological charge expectation $\braket{Q}$ can be used as an order parameter to detect the SSB of the CP symmetry.
If the CP symmetry at $\theta=\pi$ survives, $\braket{Q}_\theta$ is a continuous function of $\theta$ and vanishes at $\theta=\pi$.
On the other hand, if the CP symmetry is spontaneously broken, 
$\braket{Q}_\theta$ is discontinuous at $\theta=\pi$ indicating the first-order phase transition.
% In this case, the corresponding free energy shows the so-called multi-branched structure. %%% Witten
Therefore, to investigate the phase diagram at $\theta=\pi$, 
we need to measure $\braket{Q}_\theta$ for $\theta\to\pi$ in the thermodynamic limit, namely 
\begin{equation}
    O = \lim_{\theta\to\pi} \lim_{V \to \infty} \frac{\braket{Q}_\theta}{V}, 
\end{equation}
where $V$ is the space-time volume. 
However, the sign problem prevents us from computing $\braket{Q}_\theta$ by conventional Monte Carlo simulation, 
because the Boltzmann weight becomes complex due to $e^{i\theta Q}$.

In this work, we circumvent this problem using imaginary $\theta$ simulations.
Regarding $\theta$ as a pure imaginary number $\theta = i\tilde{\theta}$ with a new parameter $\tilde{\theta} \in \mathbb{R}$, 
the Boltzmann weight becomes a real factor $e^{-\tilde{\theta}Q}$.
The sign problem does not occur in this case, and thus various Monte Carlo techniques are applicable.
The order parameter for the original real $\theta$ is estimated by analytic continuation.

%%%%%%%%%%%%%%%%%%%%%%%%%%%%%%%%%%%%%%%%%%%%%%%%%%%%%%%%%%%%%%%%%%%%%%%%%%%%%%%%
\section{Lattice regularization and simulation method}
%%%%%%%%%%%%%%%%%%%%%%%%%%%%%%%%%%%%%%%%%%%%%%%%%%%%%%%%%%%%%%%%%%%%%%%%%%%%%%%%

For the Monte Carlo simulation of SU(3) Yang-Mills theory, we employ the renormalization group improved gauge action \cite{Iwasaki:1985we, Iwasaki:2011np}, 
\begin{equation}
    S_g[U] = -\frac{\beta}{2N} \sum_{n} \sum_{\mu\neq\nu}
    \mathrm{Tr}\left[ c_0 P_{n}^{\mu\nu} + c_1 \left( V_{n}^{\mu\nu} + H_{n}^{\mu\nu} \right) \right],
\end{equation}
where $\beta = 2N/g_0^2$ is the lattice coupling constant, $c_0 = 1 - 8 c_1$, and $c_1 = -0.331$.
$P_{n}^{\mu\nu}$, $V_{n}^{\mu\nu}$, and $H_{n}^{\mu\nu}$ represent the $1 \times 1$, $2 \times 1$, and $1 \times 2$ Wilson loops, 
respectively, at the lattice site $n$ along the $\mu$-$\nu$ direction.

As for the topological charge, we adopt the simplest clover-leaf definition \cite{DiVecchia:1981aev}, 
\begin{equation}
    Q_\mathrm{L}[U] = -\frac{1}{32\pi^{2}} \sum_{n} \frac{1}{2^{4}} \sum_{\mu,\nu,\rho,\sigma=\pm1}^{\pm4}
    \tilde{\epsilon}_{\mu\nu\rho\sigma} \mathrm{Tr} \left[P_{n}^{\mu\nu}P_{n}^{\rho\sigma}\right],
\end{equation}
where $\tilde{\epsilon}_{\mu\nu\rho\sigma}$ is an anti-symmetric tensor 
that satisfies $\tilde{\epsilon}_{(-\mu)\nu\rho\sigma} = -\epsilon_{\mu\nu\rho\sigma}$.
However, it is known that the lattice topological charge defined in this manner is subject to severe UV fluctuations of the gauge field.
The resulting topological charge is far from an integer, and the nontrivial CP symmetry at $\theta=\pi$ never appears.
Thus, we suppress the fluctuations by stout smearing \cite{Morningstar:2003gk} to recover the topological nature.
The most important advantage of stout smearing is that the method can be combined with hybrid Monte Carlo simulation.
Here we define the lattice topological charge using the smeared link variable $\mathcal{U}$, and the total action is 
\begin{equation}
    S = S_g[U] + \tilde{\theta}_\mathrm{L} \, Q_\mathrm{L} [\mathcal{U}],
\end{equation}
where $\tilde{\theta}_\mathrm{L}$ is the lattice version of the imaginary $\theta$ parameter.
The drift force of $U$ is no longer written by local link variables but is a complicated nonlocal function, 
which can be computed by reversing the smearing steps.
Therefore, the effect of the smearing is reflected not only in the measurement but also in the configuration generation.

\begin{figure}[t]
    \centering
    \includegraphics[width=0.5\linewidth]{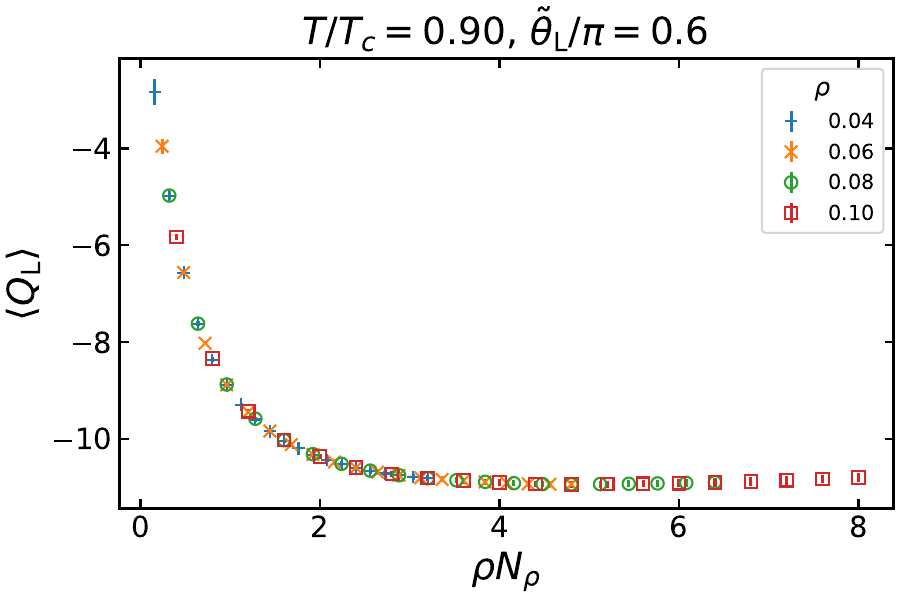}
    \caption{The topological charge expectation value is plotted against the flow time $\rho N_\rho$ for several values of $\rho$.
    The simulation was performed on the $16^3 \times 5$ lattice at $T/T_c = 0.9$ and $\tilde{\theta}_\mathrm{L}/\pi = 0.6$.}
    \label{fig_Q_flow}
\end{figure}

In the stout smearing method, there are two parameters to be tuned.
One is the smearing step size $\rho$, which should be small enough to suppress the finite-step-size effect.
The other is the number of smearing steps $N_\rho$, which should be large enough to obtain the topological charge close to integers.
Since the computational cost depends linearly on $N_\rho$, 
we set $\rho$ as large as possible so that the flow time $\rho N_\rho$ is sufficiently large with reasonable choice of $N_\rho$.
In Figure~\ref{fig_Q_flow}, the topological charge expectation value $\braket{Q_\mathrm{L}}$ is plotted against the flow time $\rho N_\rho$ 
for $\rho = 0.04$, $0.06$, $0.08$, and $0.10$.
For these choices of $\rho$, the data points of $\braket{Q_\mathrm{L}}$ collapse into a single curve.
Thus, we observe scaling behavior with respect to the flow time, indicating that the finite-step-size effect is sufficiently small.
Note that if $\rho$ is too large, the data points deviate from this curve.

\begin{figure}[t]
    \centering
    \includegraphics[width=0.8\linewidth]{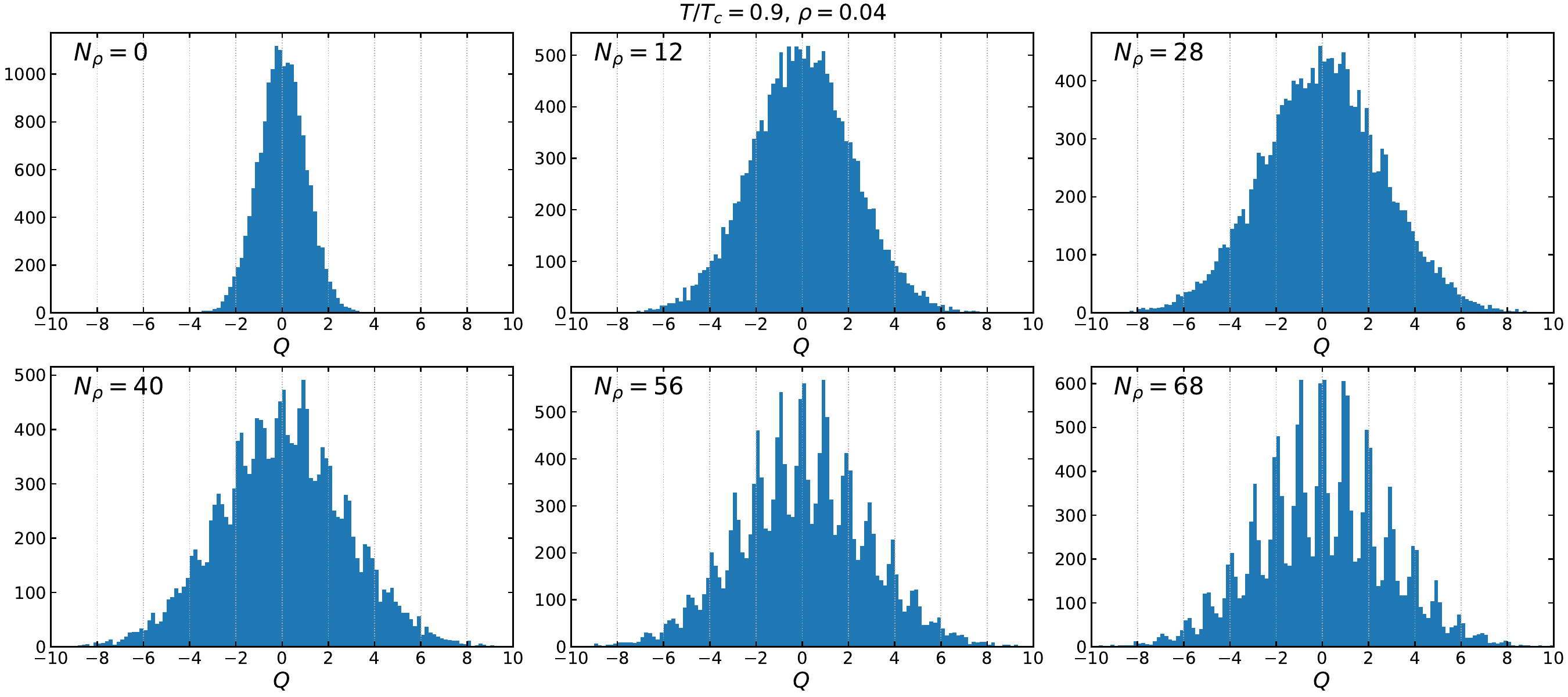}
    \caption{The histograms of the rescaled topological charge $Q=wQ_\mathrm{L}[\mathcal{U}]$ at $\theta=0$ are shown for several values of $N_\rho$.
    The simulation was performed on the $16^3 \times 5$ lattice at $T/T_c = 0.90$ and $\rho = 0.04$.}
    \label{fig_Q_dist}
\end{figure}

As the flow time $\rho N_\rho$ increases, the lattice topological charge defined with the stout smearing approaches an integer.
However, the peak positions of the topological charge distribution are slightly shifted from integers due to the lattice artifact.
Thus, we rescale the topological charge, $Q_\mathrm{L} \to wQ_\mathrm{L}$, with a parameter $w$ so that the peak positions become integers.
The optimal value of $w$ is determined by minimizing the cost function, 
\begin{equation}
    F(w)=\Braket{1-\cos(2\pi w Q_\mathrm{L}[\mathcal{U}])},
\end{equation}
where $\braket{\cdots}$ denotes the ensemble average\footnote{
Similar methods are used in Refs.~\cite{DelDebbio:2002xa, Bonati:2015sqt}, for instance.}.
The resulting $w$ depends on the temperature but is typically $w \sim 1.2$.
In Figure~\ref{fig_Q_dist}, the histograms of the topological charge are plotted against $w Q_\mathrm{L}[\mathcal{U}]$ 
for $N_\rho = 0$, $12$, $28$, $40$, $56$, and $68$ with the fixed $\rho = 0.04$.
The histograms show peaks at integer values for sufficiently large $N_\rho$.
In the following analyses, we use the rescaled topological charge $Q:= w Q_\mathrm{L}[\mathcal{U}]$ 
and the rescaled $\theta$ parameter $\theta:= \theta_\mathrm{L}/w$.

%%%%%%%%%%%%%%%%%%%%%%%%%%%%%%%%%%%%%%%%%%%%%%%%%%%%%%%%%%%%%%%%%%%%%%%%%%%%%%%%
\section{Two-dimensional version of the parallel tempering}
%%%%%%%%%%%%%%%%%%%%%%%%%%%%%%%%%%%%%%%%%%%%%%%%%%%%%%%%%%%%%%%%%%%%%%%%%%%%%%%%

To reduce the long autocorrelation of the topological charge, 
we use a two-dimensional version of the parallel tempering, which was originally proposed in Ref.~\cite{sugita2000multidimensional}.
We define a generalized ensemble which consists of non-interacting replicas as 
\begin{equation}
    Z_{\rm gen} = \prod_{i=1}^{N_\beta} \prod_{j=1}^{N_{\tilde{\theta}}} Z(\beta_i, \tilde{\theta}_j),
\end{equation}
where $N_\beta$ and $N_{\tilde{\theta}}$ are the numbers of replicas in the $\beta$ and $\tilde{\theta}$ directions, respectively.
Therefore, we have $N_\beta \times N_{\tilde{\theta}}$ replicas in total.

In order to perform the replica exchange, we choose pairs of replicas in the following way.
For the $2n$-th trial, replica exchanges are performed between pairs of replicas with the same value of $\tilde{\theta}$. 
Therefore, replicas are exchanged between $Z(\beta_{(2k-1)}, \tilde{\theta}_j)$ and $Z(\beta_{(2k)}, \tilde{\theta}_j)$ at each $\tilde{\theta}$, 
where $k$ runs from 1 to $N_\beta / 2$.
On the contrary, for the $(2n+1)$-th trial, replica exchanges are performed between pairs of replicas with the same values of $\beta$. 
Therefore, replicas are excanged between $Z(\beta_{i}, \tilde{\theta}_{(2l-1)})$ and $Z(\beta_{i}, \tilde{\theta}_{(2l)})$ at each $\beta$, 
where $l$ runs from 1 to $N_{\tilde{\theta}} / 2$.
The replica exchange is performed under the probability 
\begin{equation}
    P=\min (1,e^{\Delta S})
\end{equation}
with
\begin{equation}
\begin{split}
    \Delta S &= S[\beta,\tilde{\theta};U^\prime] - S[\beta,\tilde{\theta};U] 
                    + S[\beta^\prime,\tilde{\theta}^\prime;U] - S[\beta^\prime,\tilde{\theta}^\prime;U^\prime]\\
            & = (\beta -\beta^\prime) \left(S_g[U]/\beta - S_g[U^\prime]/\beta^\prime\right) - (\tilde{\theta} -\tilde{\theta}^\prime)(Q[U] - Q[U^\prime]),
\end{split}
\end{equation}
where configurations $U$ and $U^\prime$ are generated at $(\beta,\tilde{\theta})$ and $(\beta^\prime,\tilde{\theta}^\prime)$, respectively.
For the choice of parameters, we choose a relatively small step for $\beta$ and a relatively large step for $\tilde{\theta}$ 
because $S_g \propto V$ while $Q \lesssim \mathcal{O}(10)$ in our setup.

\begin{figure}
    \centering
    \includegraphics[width=0.5\linewidth]{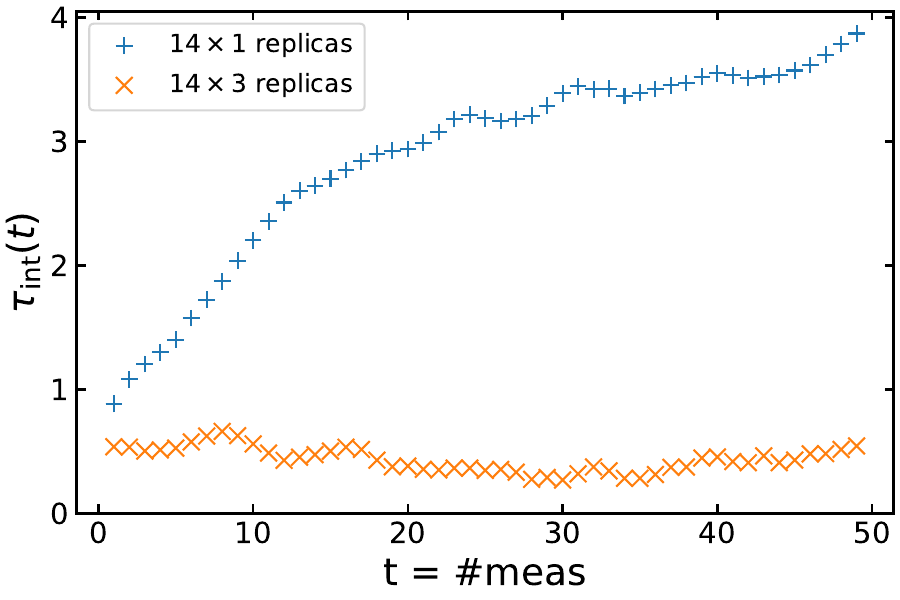}
    \caption{The integrated autocorrelation time at $(T/T_c,~ \tilde{\theta}_\mathrm{L}) = (1.0,~ 0.5)$ 
    obtained by the 1d and 2d versions of parallel tempering.}
    \label{fig:parallel_tempering}
\end{figure}

To evaluate the autocorrelation with the 1-dimensional and 2-dimensional parallel tempering, we perform simulations with 
\begin{align*}
    & T/T_\mathrm{dec}(0) = \{0.93, 0.94,...,1.06\}\ \text{ and }\ \tilde{\theta}_\mathrm{L}/\pi = 0.5 && \text{for the 1d version},\\
    & T/T_\mathrm{dec}(0) = \{0.93, 0.94,...,1.06\}\ \text{ and }\ \tilde{\theta}_\mathrm{L}/\pi = \{0.4, 0.5, 0.6\} && \text{for the 2d version}.
\end{align*}
In Figure~\ref{fig:parallel_tempering}, the integrated autocorrelation time $\tau_{int}$ of topological charge 
is plotted aginst Monte Carlo time $t$ at $(T/T_c,~ \tilde{\theta}_\mathrm{L}) = (1.0,~ 0.5)$.
We find that $\tau_{\rm int}$ for the 2d version is 6 times smaller than that for the 1d version.
Considering that the 2d version has 3 times more parameters than that of 1d version, the 2d version is 2 times more efficient than the 1d version.

%%%%%%%%%%%%%%%%%%%%%%%%%%%%%%%%%%%%%%%%%%%%%%%%%%%%%%%%%%%%%%%%%%%%%%%%%%%%%%%%
\section{Numerical results}
%%%%%%%%%%%%%%%%%%%%%%%%%%%%%%%%%%%%%%%%%%%%%%%%%%%%%%%%%%%%%%%%%%%%%%%%%%%%%%%%

In this section, we show our numerical results obtained by hybrid Monte Carlo simulation with the 2-dimensional parallel tempering.
The lattice size is fixed to $16^3 \times 5$, 
and the relation between the lattice coupling $\beta$ and temperature $T/T_c$ is determined by the scaling function in Ref.~\cite{CP-PACS:1999eop}.
Here $T_c$ denotes the deconfining temperature in the continuum limit.
The smearing step size and the number of steps are set to $\rho = 0.04$ and $N_\rho = 80$, 
which corresponds to the flow time $\rho N_\rho = 3.2$ in the scaling region of Figure~\ref{fig_Q_flow}.

%%%%%%%%%%%%%%%%%%%%%%%%%%%%%%%%%%%%%%%%
\subsection{CP symmetry at \texorpdfstring{$\theta=\pi$}{theta = pi}}
%%%%%%%%%%%%%%%%%%%%%%%%%%%%%%%%%%%%%%%%

\begin{figure}
    \centering
    \includegraphics[width=0.65\linewidth]{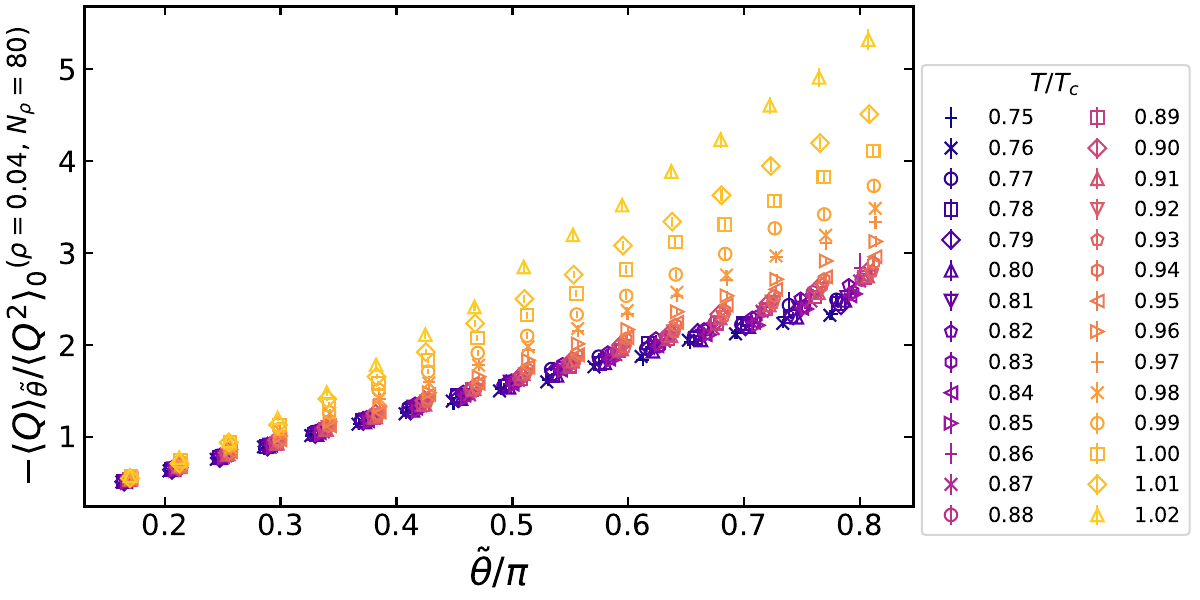}
    \caption{The normalized topological charge expectation value $-\braket{Q}_{\tilde{\theta}}/\braket{Q^2}_0$ 
    is plotted against $\tilde{\theta}/\pi$ for various temperatures $0.75 \le T/T_c \le 1.02$.
    The simulation was performed on the $16^3 \times 5$ lattice with the smearing parameters $\rho = 0.04$ and $N_\rho = 80$.}
    \label{fig_Q_itheta}
\end{figure}

First, we present the imaginary $\theta$ dependence of the topological charge expectation value $\braket{Q}_{\tilde{\theta}}$.
In Figure~\ref{fig_Q_itheta}, $\braket{Q}_{\tilde{\theta}}$ is plotted against $\tilde{\theta}/\pi$ 
for various temperatures $0.75 \le T/T_c \le 1.02$ with normalization by the variance $\braket{Q^2}_0$ at $\tilde{\theta}=0$.
At low temperatures, $\braket{Q}_{\tilde{\theta}}$ depends almost linearly on $\tilde{\theta}$.
In this case, analytic continuation to real $\theta$ yields $\braket{Q}_{\theta} \propto i\theta$ for $|\theta|<\pi$, 
indicating that the CP symmetry at $\theta=\pi$ is broken.
Indeed, the system at low temperatures is similar to the large-$N$ limit of SU($N$) Yang-Mills theory, 
which exhibits the linear behavior at low temperatures \cite{Witten:1980sp, Witten:1998uka}.

On the other hand, as the temperature increases, a nonlinear behavior gradually appears.
At sufficiently high temperatures, the system is expected to be approximated by the dilute instanton gas \cite{Gross:1980br, Weiss:1980rj}, 
where the CP symmetry at $\theta=\pi$ is recovered.
In the instanton gas model, the topological charge expectation value is a smooth function, $\braket{Q}_{\theta} \propto i\sin\theta$, 
and the corresponding behavior on the imaginary $\theta$ side is $\braket{Q}_{\tilde{\theta}} \propto \sinh\tilde{\theta}$.
The growth of nonlinearity in the result is consistent with this prediction.

\begin{figure}
    \centering
    \includegraphics[width=0.85\linewidth]{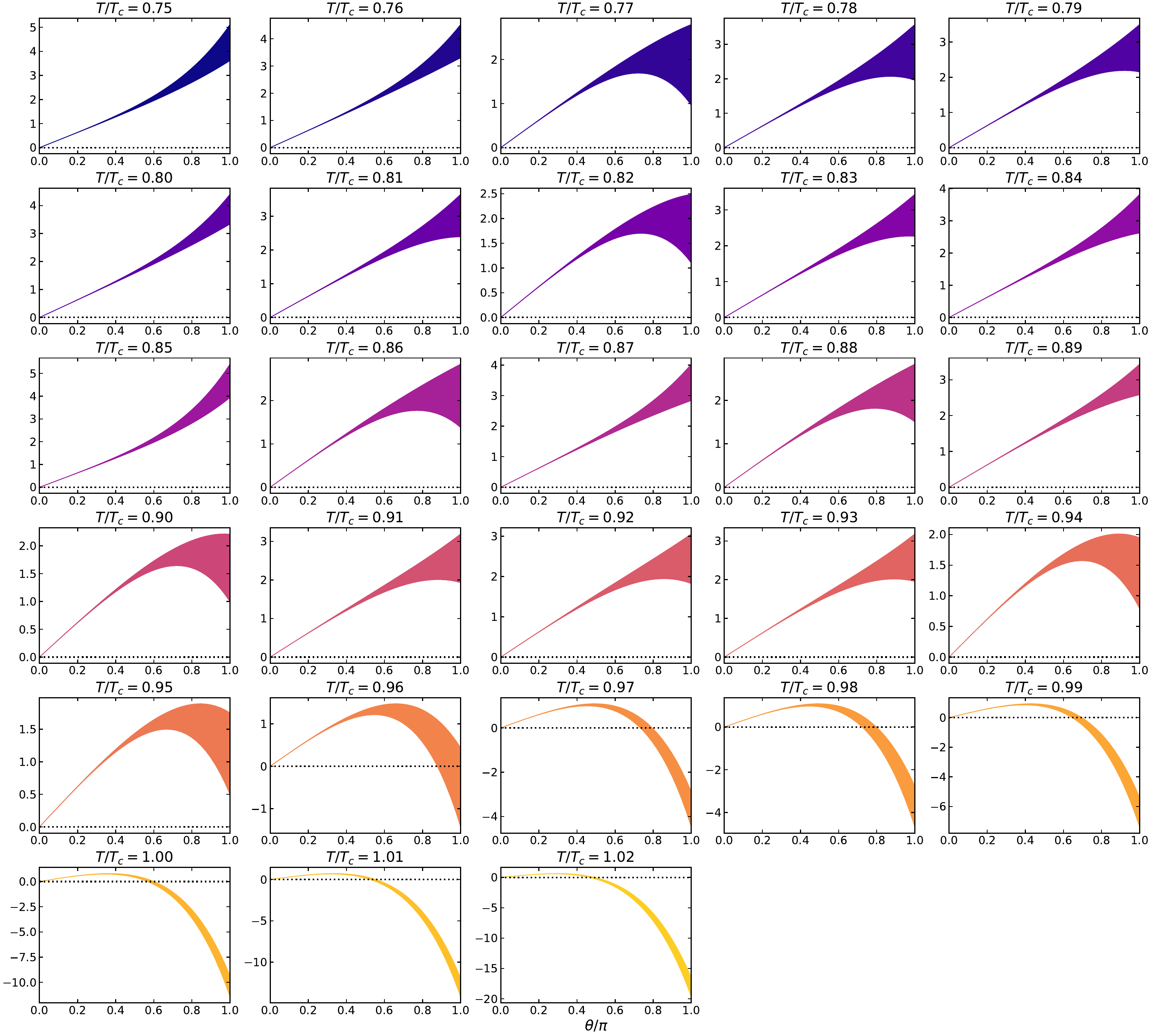}
    \caption{The fitting curves of the data points in Figure~\ref{fig_Q_itheta} are analytically continued to real $\theta$, 
    and plotted against $\theta/\pi$ with the error bands.}
    \label{fig_analytic_cont}
\end{figure}

To perform analytic continuation of the numerical results from imaginary $\theta$ to real $\theta$, we consider a fitting ansatz, 
\begin{equation}
    \frac{\braket{Q}_{\tilde{\theta}}}{\braket{Q^2}_0} 
    = -\left( \tilde{\theta} + \frac{a_3}{6} \tilde{\theta}^3 + \frac{a_5}{120} \tilde{\theta}^5 \right), 
    \label{eq_fit_ansatz}
\end{equation}
which is a straightforward generalization of the linear behavior of the large-$N$ Yang-Mills theory.
The higher-order terms are responsible for the nonlinear correction needed to approach the $\sinh$ behavior.
We fit the data points in Figure~\ref{fig_Q_itheta} with this ansatz and apply analytic continuation, 
\begin{equation}
    \frac{\braket{Q}_{\theta}}{\braket{Q^2}_0} 
    = i\left( \theta - \frac{a_3}{6} \theta^3 + \frac{a_5}{120} \theta^5 \right).
\end{equation}
Note that the coefficient of the first term is fixed to 1 by the definition of the topological susceptibility $\chi_0$ at $\theta=0$, 
\begin{equation}
    \chi_0 V = \braket{Q^2}_0 = \left. i\frac{\partial \braket{Q}_{\theta}}{\partial \theta} \right|_{\theta=0}.
\end{equation}
The resulting fitting curves after analytic continuation are shown in Figure~\ref{fig_analytic_cont} with the error bands, 
where the horizontal axis is $\theta/\pi$.
The fitting curve crossing zero at $\theta=\pi$ indicates that the CP symmetry is recovered.
Thus, the estimated CP restoration temperature is $T_{\mathrm{CP}}/T_c \sim 0.96$ for the current lattice size.
Note that the fitting curves for $T/T_c > 0.96$ at $\theta=\pi$ go down to negative values, 
which is thought to be due to the low-order approximation of the $\sinh$ function in the fitting ansatz~\ref{eq_fit_ansatz}.

%%%%%%%%%%%%%%%%%%%%%%%%%%%%%%%%%%%%%%%%
\subsection{Deconfining temperature for nonzero \texorpdfstring{$\theta$}{theta}}
%%%%%%%%%%%%%%%%%%%%%%%%%%%%%%%%%%%%%%%%

Next, we focus on the $\theta$ dependence of the deconfining temperature.
To determine the deconfining temperature $T_\mathrm{dec}(\theta)$ at $\theta \neq 0$ for a finite lattice, 
we measure the Polyakov loop susceptibility. 
Since the phase transition is of the first order, the susceptibility has a peak at the critical point at finite volume, 
and it diverges in the infinite volume limit.
Thus, we can identify the critical point as the peak position of the susceptibility.

\begin{figure}
    \centering
    \includegraphics[width=0.6\linewidth]{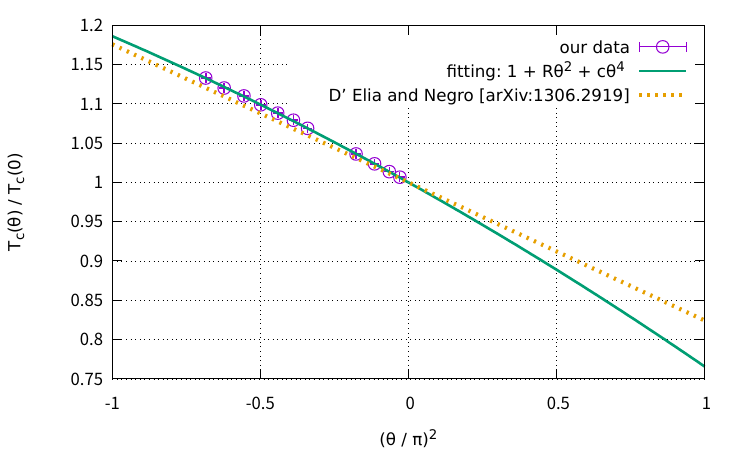}
    \caption{The deconfining temperature $T_\mathrm{dec}(\theta)/T_c$ for the finite lattice is plotted against $(\theta/\pi)^2$.
    The solid line represents the fitting result with Eq.~(\ref{eq_fit_Tdec}).
    The dotted curve is the result in Ref.~\cite{DElia:2013uaf}.}
    \label{fig_Tdec}
\end{figure}

We perform simulations with the imaginary $\theta$ for $T/T_c \in [0.997,\, 1.156]$ and determine the deconfining temperature.
In Figure~\ref{fig_Tdec}, the result of $T_\mathrm{dec}(\theta)/T_c$ is plotted against $(\theta/\pi)^2$.
We fit the data points with the ansatz 
\begin{equation}
    T_\mathrm{dec}(\theta)/T_c = 1 - R \theta^2 + c \theta^4,
    \label{eq_fit_Tdec}
\end{equation}
and obtain $R = 0.0213(1)$ and $c = -0.00025(3)$.
The fitting result is shown in Figure~\ref{fig_Tdec}, 
where the left and right regions of the plot correspond to the imaginary and real $\theta$, respectively, 
and they are connected by analytic continuation.
Our result at fixed volume and at fixed lattice spacing is qualitatively the same as the result 
in the continuum and infinite-volume limits of the previous work~\cite{DElia:2013uaf}.
Assuming that the simple polynomial ansatz~(\ref{eq_fit_Tdec}) is valid at $\theta=\pi$, 
the estimated deconfinging temperature there is around $0.75 \le T_\mathrm{dec}(\pi)/T_c \le 0.8$.

%%%%%%%%%%%%%%%%%%%%%%%%%%%%%%%%%%%%%%%%
\section{Summary}
%%%%%%%%%%%%%%%%%%%%%%%%%%%%%%%%%%%%%%%%

We investigated the phase diagram of 4D SU(3) Yang-Mills theory with the $\theta$ term, focusing on the CP symmetry at $\theta=\pi$.
Since the ordinary Monte Carlo simulation with a nonzero $\theta$ term suffers from the sign problem, we employed the model with imaginary $\theta$.
The topological nature of the gauge field on the lattice is recovered by the stout smearing method, 
which is incorporated in the hybrid Monte Carlo simulation with the 2-dimensional parallel tempering.
We compute the topological charge expectation value $\braket{Q}_{\tilde{\theta}}$, namely the order parameter of CP symmetry, 
at imaginary $\theta$, and estimate its behavior $\braket{Q}_{\theta}$ at real $\theta$ by analytic continuation.
At low temperatures, we observed a linear behavior of $\braket{Q}_{\theta}$, which corresponds to the CP broken phase.
The nonlinear behavior grows with the temperature, and we get $\braket{Q}_{\theta=\pi} \sim 0$ at $T/T_c = 0.96$, indicating the restoration of CP symmetry.
On the other hand, the simple estimation of the deconfining temperature at $\theta=\pi$ resulted in $0.75 \le T_\mathrm{dec}(\pi)/T_c \le 0.8$.
Therefore, the current result, $T_{\mathrm{CP}} \sim 0.96 T_c > T_\mathrm{dec}(\pi)$, 
for the $16^3 \times 5$ finite lattice suggests a phase diagram similar to our previous study on 4D SU(2) Yang-Mills theory~\cite{Hirasawa:2024fjt}.
Note that the finite volume effect in this result has not been estimated so far, which is an important future direction of this project.
The new computations with larger lattice sizes are ongoing, and we hope to report the result of the large-volume limit in the near future.

%%%%%%%%%%%%%%%%%%%%%%%%%%%%%%%%%%%%%%%%
\acknowledgments
%%%%%%%%%%%%%%%%%%%%%%%%%%%%%%%%%%%%%%%%

We thank Masazumi Honda, Yuta Ito, and Yuya Tanizaki for their valuable discussions and comments.
This research was conducted using Yukawa-21 at YITP in Kyoto University, 
Supermicro ARS-111GL-DNHR-LCC, and FUJITSU Server PRIMERGY CX2550 M7 (Miyabi) at Joint Center for Advanced High Performance Computing (JCAHPC).
% This research used the computational resources of Yukawa-21 at YITP, Kyoto University, 
% and Miyabi, provided by the Multidisciplinary Cooperative Research Program in the Center for Computational Sciences, University of Tsukuba.
A.~M. is supported by JST-CREST No.~JPMJCR24I1.
A.~Y. is supported by JST-CREST No.~JPMJCR24I3 and JSPS Grant-in-Aid for Transformative Research Areas (A) JP21H05191.

\bibliographystyle{JHEP}
\bibliography{ref}

\end{document}